\documentclass[letterpaper,twocolumn,10pt]{article}

\usepackage[table]{xcolor}
\usepackage{usenix,epsfig}
\usepackage[utf8]{inputenc}
\usepackage{tikz}
\usetikzlibrary{calc}
\usetikzlibrary{positioning}
\usepackage{amsmath}
\usepackage{subfigure}
\usepackage{tabularx}
\usepackage{multirow}
\usepackage{hyperref}
\usepackage{balance}
\usepackage{xspace}
\usepackage{inconsolata}
\usepackage{balance}
\usepackage[backend=bibtex,backref=true]{biblatex}
\usepackage[T1]{fontenc}
\usepackage[scaled=0.8]{beramono}
\usepackage{enumitem} 

\bibliography{paper}

\definecolor{darkblue}{rgb}{0,0,0.4}
\hypersetup{
	pdftitle={Identifying and characterizing Sybils in the Tor network},
	pdfauthor={Philipp Winter, Roya Ensafi, Karsten Loesing, and Nick Feamster},
	colorlinks=true,
	urlcolor=darkblue,
	linkcolor=darkblue,
	citecolor=darkblue
}

\newcommand{\sys}{sybilhunter\xspace}
\newcommand{\Sys}{Sybilhunter\xspace}

\begin{document}

\title{\Large \bf Identifying and characterizing Sybils in the Tor network}

\author{
{\rm Philipp Winter}\\
Princeton \& Karlstad \\ University
\and
{\rm Roya Ensafi}\\
Princeton University
\and
{\rm Karsten Loesing}\\
The Tor Project
\and
{\rm Nick Feamster}\\
Princeton University
}

\maketitle

\subsection*{Abstract}
Being a volunteer-run, distributed anonymity network, Tor is vulnerable to Sybil
attacks.  Little is known about real-world Sybils in the Tor network, and we
lack practical tools and methods to expose Sybil attacks.
In this work, we develop \emph{\sys}, the first system for detecting Sybil
relays based on their \emph{appearance}, such as configuration; and
\emph{behavior}, such as uptime sequences.  We used \sys's diverse analysis
techniques to analyze nine years of archived Tor network data, providing us with
new insights into the operation of real-world attackers.  Our findings include
diverse Sybils, ranging from botnets, to academic research, and relays that
hijack Bitcoin transactions.
Our work shows that existing Sybil defenses do not apply to Tor, it delivers
insights into real-world attacks, and provides practical tools to uncover and
characterize Sybils, making the network safer for its users.

\section{Introduction}\label{sec:introduction}
In a Sybil attack, an attacker controls many virtual identities to obtain
disproportionately large influence in a network.  These attacks take many
shapes, such as sockpuppets hijacking online discourse~\cite{Thomas2012a}; the
manipulation of BitTorrent's distributed hash table~\cite{Wang2012a}; and, most
relevant to our work, relays in the Tor network that seek to deanonymize
users~\cite{cmucert}.  In addition to coining the term ``Sybil,'' Douceur showed
that practical Sybil defenses are challenging, arguing that Sybil attacks are
always possible without a central authority~\cite{Douceur2002a}.  In this work,
we focus on Sybils in Tor---relays that are controlled by a single operator.
But what harm can Sybils do?

The effectiveness of many attacks on Tor depends on how large a fraction of the
network's traffic---the consensus weight---an attacker can observe.  As the
attacker's consensus weight grows, the following attacks become easier.

\begin{description}
	\item[Exit traffic tampering:] A Tor user's traffic traverses exit relays,
		the last hop in a Tor circuit, when leaving the Tor network.
		Controlling exit relays, an attacker can sniff traffic to collect
		unencrypted credentials, break into TLS-protected connections, or inject
		malicious content~\cite{Winter2014a}.
	\item[Website fingerprinting:] Tor's encryption prevents guard relays (the
		first hop in a Tor circuit) from learning their user's online activity.
		Ignoring the encrypted payload, an attacker can still take advantage of
		flow information such as packet lengths and timings to infer what web
		site her users are connecting to~\cite{Juarez2014a}.
	\item[Bridge address harvesting:] Users behind censorship firewalls use
		private Tor relays (``bridges'') as hidden stepping stones into the Tor
		network.  It is important that censors cannot obtain all bridge
		addresses, which is why bridge distribution is rate-limited.  However,
		an attacker can harvest bridge addresses by running a middle relay and
		looking for incoming connections that do not originate from any of the
		publicly known guard relays~\cite{Ling2015b}.
	\item[End-to-end correlation:] By running both entry guards and exit relays,
		an attacker can use timing analysis to link a Tor user's identity to her
		activity, e.g., learn that \emph{Alice} is visiting \emph{Facebook}.
		For this attack to work, an attacker must run at least two Tor relays,
		or be able to eavesdrop on at least two networks~\cite{Johnson2013a}.
\end{description}

Configuring a relay to forward more traffic allows an attacker to increase her
consensus weight.  However, the capacity of a single relay is limited by its
link bandwidth and, because of the computational cost of cryptography, by CPU.
Ultimately, increasing consensus weight requires an adversary to add relays to
the network; we call these additional relays Sybils.

In addition to the above attacks, an adversary needs Sybil relays to manipulate
onion services, which are TCP servers whose IP address is hidden by Tor.  In the
current onion service protocol, six Sybil relays are sufficient to take offline
an onion service because of a weakness in the design of the distributed hash
table (DHT) that powers onion services~\cite{Biryukov2013a}.  Finally, instead
of being a direct means to an end, Sybil relays can be a \emph{side effect} of
another issue.  In Section~\ref{sec:sybil_groups}, we provide evidence for what
appears to be botnets whose zombies are running Tor relays, perhaps because of a
misguided attempt to help the Tor network grow.

Motivated by the lack of practical Sybil detection tools, we design and
implement heuristics, leveraging that Sybils (\emph{i}) frequently go online and
offline simultaneously, (\emph{ii}) share similarities in their configuration,
and (\emph{iii}) may change their identity fingerprint---a relay's fingerprint
is the hash over its public key---frequently, to manipulate Tor's DHT.  We
implemented these heuristics in a tool, \sys, whose development required a major
engineering effort because we had to process 100 GiB of data and millions of
files.  We used \sys to analyze archived network data, dating back to 2007, to
discover past attacks and anomalies.  Finally, we characterize the Sybil groups
we discovered.  To sum up, we make the following key contributions:
\begin{itemize}
	\item We design and implement \sys, a tool to analyze past and future Tor
		network data.  While we designed it specifically for the use in Tor, our
		techniques are general in nature and can easily be applied to other
		distributed systems such as I2P~\cite{i2p}.
	\item We expose and characterize Sybil groups, and publish our findings as
		datasets to stimulate future research.\footnote{The dataset is online at
			\url{https://nymity.ch/sybilhunting/}.} We find that Sybils run MitM
		attacks, DoS attacks, and are used for research projects.
\end{itemize}

The rest of this paper is structured as follows.  We begin by discussing related
work in Section~\ref{sec:related_work} and give some background on Tor in
Section~\ref{sec:background}.  Section~\ref{sec:design} presents the design of
our analysis tools, which is then followed by experimental results in
Section~\ref{sec:results}.  We discuss our results in
Section~\ref{sec:discussion} and conclude the paper in
Section~\ref{sec:conclusion}.

\section{Related work}\label{sec:related_work}
In his seminal 2002 paper, Douceur showed that only a \emph{central authority}
that verifies new nodes as they join the distributed system is guaranteed to
prevent Sybils~\cite{Douceur2002a}.  This approach conflicts with Tor's design
philosophy that seeks to distribute trust and eliminate central points of
control.  In addition, a major factor contributing to Tor's network growth is
the low barrier of entry, allowing operators to set up relays both quickly and
anonymously.  An identity-verifying authority would raise that barrier, alienate
privacy-conscious relay operators, and impede Tor's growth.  Barring a central
authority, researchers have proposed techniques that leverage a resource that is
difficult for an attacker to scale.  Two categories of Sybil-resistant schemes
turned out to be particularly popular, schemes that build on \emph{social
constraints} and schemes that build on \emph{computational constraints}.  For a
broad overview of alternative Sybil defenses, refer to Levine et
al.~\cite{Levine2006a}.

Social constraints rely on the assumption that it is difficult for an attacker
to form trust relationships with honest users, e.g., befriend many unknown
people on online social networks.  Past work leveraged this assumption in
systems such as SybilGuard~\cite{Yu2006a}, SybilLimit~\cite{Yu2008a}, and
SybilInfer~\cite{Danezis2009a}.  Unfortunately, social graph-based defenses
do not work in our setting because there is no existing trust relationship
between relay operators.\footnote{Relay operators can express in their
configuration that their relays are run by the same operator, but this
denotes an \emph{intra}-person and not an \emph{inter}-person trust
relationship.} Note that we could create such a relationship by, e.g., linking
relays to their operator's social networking account, or by creating a ``relay
operator web of trust,'' but again, we believe that such an effort would
alienate relay operators and receive limited adoption.

Orthogonal to social constraints, computational resource constraints guarantee
that an attacker that seeks to operate 100 Sybils needs 100 times the
computational resources she would have needed for a single virtual identity.
Both Borisov~\cite{Borisov2006a} and Li et al.~\cite{Li2012a} used computational
puzzles for that purpose.  Computational constraints work well in distributed
systems where the cost of joining the network is low.  For example, a
lightweight client is enough to use BitTorrent, allowing even low-end consumer
devices to participate.  However, this is not the case in Tor because relay
operations require constant use of bandwidth and CPU.  Unlike in many other
distributed systems, it is impossible to run 100 Tor relays while not spending
the resources for 100 relays.  Computational constraints are inherent to running
a relay.

There has also been progress outside of academic research; namely, The Tor
Project has incorporated a number of both implicit and explicit Sybil defenses
that are in place as of February 2016.  First, directory authorities---the
``gatekeepers'' of the Tor network---accept at most two relays per IP address to
prevent low-resource Sybil attacks~\cite{Bauer2007a,Bauer2007b}.  Similarly,
Tor's path selection algorithm states that Tor clients never select two relays
in the same /16 network~\cite{path-spec}.  Second, directory authorities
automatically assign flags to relays, indicating their status and quality of
service.  The Tor Project has recently increased the minimal time until relays
obtain the \texttt{Stable} flag (seven days) and the \texttt{HSDir} flag (96
hours).  This change increases the cost of Sybil attacks and gives Tor
developers more time to discover and block suspicious relays before they get in
a position to run an attack.  Finally, the operation of a Tor relay causes
recurring costs---most notably bandwidth and electricity---which can further
restrain an adversary.

In summary, we believe that existing Sybil defenses do not work well when
applied to the Tor network; its distinctive features call for customized
solutions that consider the nature of Tor relays.

\section{Background}\label{sec:background}
We now provide necessary background on the Tor network~\cite{Dingledine2004a}.
Tor consists of several thousand volunteer-run relays that are summarized in the
\emph{network consensus} that is voted on and published every hour by eight
distributed \emph{directory authorities}.  The authorities assign a variety of
flags to relays:

\begin{description}[noitemsep]
	\item[Valid:] The relay is valid, i.e., not known to be broken.
	\item[HSDir:] The relay is an onion service directory, i.e., it participates
		in the DHT that powers Tor onion services.
	\item[Exit:] The relay is an exit relay.
	\item[BadExit:] The relay is an exit relay but is either misconfigured or
		malicious, and should therefore not be used by Tor clients.
	\item[Stable:] Relays are stable if their mean time between failure is at
		least the median of all relays, or at least seven days.
	\item[Guard:] Guard relays are the rarely-changing first hop for Tor clients.
	\item[Running:] A relay is running if the directory authorities could
		connect to it in the last 45 minutes.
\end{description}

Tor relays are uniquely identified by their \emph{fingerprint}, a Base32-encoded
and truncated SHA-1 hash over their public key.  Operators can further assign a
\emph{nickname} to their Tor relays, which is a string that identifies a relay
(albeit not uniquely) and is easier to remember than its pseudo-random
fingerprint.  Exit relays have an \emph{exit policy}---a list of IP addresses
and ports that the relay allows connections to.  Finally, operators that run
more than one relay are encouraged to configure their relays to be part of a
\emph{relay family}.  Families are used to express that a set of relays is
controlled by a single party.  Tor clients never use more than one family member
in their path to prevent correlation attacks.  As of February 2016, there are
approximately 400 relay families in all 7,000 relays.

\section{Data and design}\label{sec:design}
We define Sybils in the Tor network as two or more relays that are controlled by
a single person or group of people.  Sybils per se do not have to be malicious;
a relay operator could simply have forgotten to configure her relays as a relay
family.  Such Sybils are no threat to the Tor network, which is why we refer to
them as \emph{benign Sybils}.  What we are interested in is \emph{malicious
Sybils} whose purpose is to deanonymize or otherwise harm Tor users.

\begin{figure}[t]
	\centering
	\includegraphics[width=\linewidth]{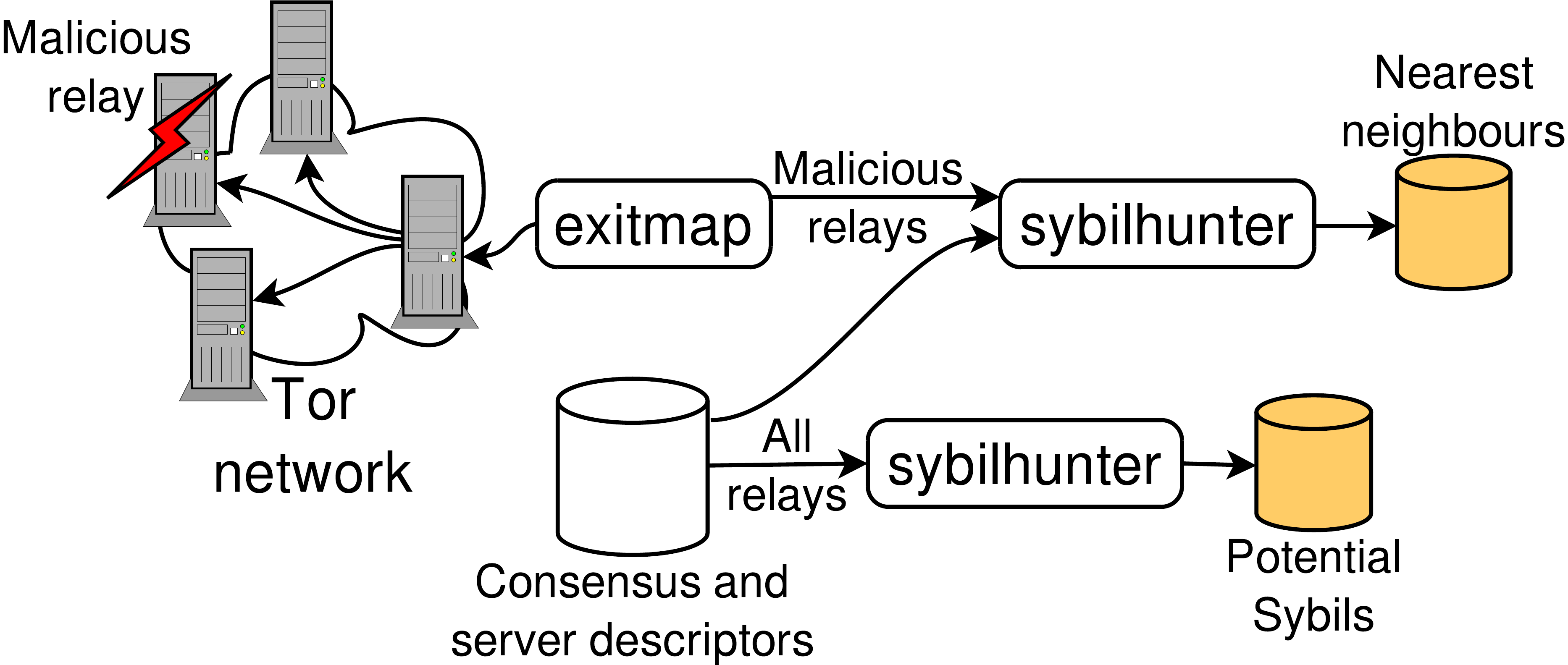}
	\caption{\Sys's architecture.  Two datasets serve as input to
		\sys; consensuses and server descriptors, and malicious
		relays gathered with exitmap~\cite{Winter2014a}.}
	\label{fig:system}
\end{figure}

We draw on two datasets---one publicly available and one created by us---to
uncover malicious Sybils.  Our detection methods are implemented in a tool,
\sys, which takes as input our two datasets and then attempts to expose Sybil
groups, as illustrated in Figure~\ref{fig:system}.  \Sys is implemented in Go
and consists of 3,300 lines of code.

\subsection{Datasets}
\label{sec:datasets}
Figure~\ref{fig:system} shows how we use our two datasets.  Archived consensuses
and router descriptors (in short: descriptors) allow us to (\emph{i}) restore
past states of the Tor network, which \sys mines for Sybil groups, and to
(\emph{ii}) find ``partners in crime'' of malicious exit relays that we
discovered by running exitmap, a scanner for Tor exit relays that is presented
below.

\subsubsection{Consensuses and router descriptors}
The consensus and descriptor dataset is publicly available on
CollecTor~\cite{collector}, an archiving service that is run by The Tor Project.
Some of the archived data dates back to 2004, allowing us to restore arbitrary
Tor network configurations from the last ten years.  Not all of CollecTor's
archived data is relevant to our hunt for Sybils, however, which is why we only
analyze the following two:

\begin{figure}[t]
\centering
\begin{tikzpicture}[node distance=0 cm,outer sep = 0pt,inner sep = 2pt]

	\tikzstyle{every node}=[font=\footnotesize]

	\tikzset{field/.style={align=center,shape=rectangle,
		minimum height=4mm,minimum width=23mm,fill=gray!20}}

	\tikzset{blankfield/.style={align=center,shape=rectangle,
		minimum height=4mm,minimum width=23mm}}

	\usetikzlibrary{positioning}

	\node (consensus) [blankfield] {\textbf{Consensus}};

	\node (routerstatus) [blankfield,below=of consensus] {\textbf{Router status}};
	\node (desc) [field,below=of routerstatus] {Descriptor pointer};
	\node (nickname) [field,below=of desc] {Nickname};
	\node (fingerprint) [field,below=of nickname] {Fingerprint};
	\node (publication) [field,below=of fingerprint] {Publication};
	\node (address) [field,below=of publication] {Address and ports};
	\node (flags) [field,below=of address] {Flags};
	\node (version) [field,below=of flags] {Version};
	\node (bandwidth) [field,below=of version] {Bandwidth};
	\node (policy) [field,below=of bandwidth] {Exit policy};
	\node [below=of policy] {\textbf{\dots}};

	\draw [draw=black] (desc.north west) rectangle (policy.south east);

	\draw [draw=black] ($(desc.north west) + (-2mm,4.5mm)$) rectangle ($(policy.south east) + (2mm,-4.5mm)$);

	\node (dummy) at (4cm,0) [blankfield] {};
	\node (descriptor) at (4cm,0) [blankfield,below=of dummy] {\textbf{Router descriptor}};
	\node (descaddr) at (4cm,0) [field,below=of descriptor] {Address and ports};
	\node (descplatform) at (4cm,0) [field,below=of descaddr] {Platform};
	\node (descprotocols) at (4cm,0) [field,below=of descplatform] {Protocols};
	\node (descpublished) at (4cm,0) [field,below=of descprotocols] {Published};
	\node (descfingerprint) at (4cm,0) [field,below=of descpublished] {Fingerprint};
	\node (descuptime) at (4cm,0) [field,below=of descfingerprint] {Uptime};
	\node (descbandwidth) at (4cm,0) [field,below=of descuptime] {Bandwidth};
	\node (descsignature) at (4cm,0) [field,below=of descbandwidth] {Signature};

	\draw [draw=black] (descaddr.north west) rectangle (descsignature.south east);

	\draw [-latex,thick] (desc) -- (descaddr);
\end{tikzpicture}
\caption{Our primary dataset contains consensuses and router descriptors.}
\label{fig:datasets}
\end{figure}
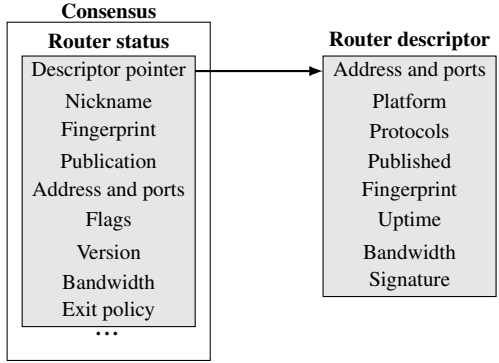

\paragraph{Descriptors} Tor relays and bridges periodically upload router
descriptors, which capture their configuration, to directory authorities.
Figure~\ref{fig:datasets} shows an example in the box to the right.  Relays
upload their descriptors no later than every 18 hours, or sooner, depending on
certain conditions.  Note that some information in router descriptors is not
verified by directory authorities.  Therefore, relays can spoof information such
as their operating system, Tor version, and uptime.

\paragraph{Consensuses} Each hour, the nine directory authorities vote on their
view of all Tor relays that are currently online.  The vote produces the
consensus, an authoritative list that comprises all running Tor relays,
represented as a set of router statuses.  Each router status in the consensus
contains basic information about Tor relays such as their bandwidth, flags, and
exit policy.  It also contains a pointer to the relay's descriptor, as shown in
Figure~\ref{fig:datasets}.  As of February 2016, consensuses contain
approximately 7,000 router statuses, i.e., each hour, 7,000 router statuses are
published, and archived, by CollecTor.

Table~\ref{tab:collector-dataset} gives an overview of the size of our consensus
and descriptor archives.  We found it challenging to repeatedly process these
millions of files, amounting to more than 100 GiB of uncompressed data.  In our
first processing attempt, we used the Python parsing library Stem~\cite{stem},
which is maintained by The Tor Project.  The data volume turned out to be
difficult to handle for Stem because of Python's interpreted and dynamic nature.
To process our dataset more efficiently, we implemented a custom parser in
Go~\cite{zoossh}.

\begin{table}[t]
\small
\centering
\begin{tabular}{l c c c}
\hline
\textbf{Dataset} & \textbf{\# of files} & \textbf{Size} & \textbf{Time span} \\
\hline
Consensuses & 72,061 & 51 GiB & 10/2007--01/2016 \\
Descriptors & 34,789,777 & 52 GiB & 12/2005--01/2016 \\
\hline
\end{tabular}
\caption{An overview of our primary dataset; consensuses and server descriptors
since 2007 and 2005, respectively.}
\label{tab:collector-dataset}
\end{table}

\subsubsection{Malicious exit relays}
In addition to our publicly available and primary dataset, we collected
malicious exit relays over 18 months.  We call exit relays malicious if they
modify forwarded traffic in bad faith, e.g., to run man-in-the-middle attacks.
We add these relays to our dataset because they frequently \emph{surface in
groups}, as malicious Sybils, because an attacker runs the same attack on
several, physically distinct exit relays.  Winter et al.'s work~\cite[\S
5.2]{Winter2014a} further showed that attackers make an effort to stay under the
radar, which is why we cannot only rely on active probing to find such relays.
We also seek to find potential ``partners in crime'' of each newly discovered
malicious relay, which we discuss in Section~\ref{sec:nearest-neighbor}.

We exposed malicious exit relays using Winter et al.'s exitmap
tool~\cite{Winter2014a}.  Exitmap is a Python-based scanning framework for Tor
exit relays.  Exitmap modules perform a network task that can then be run over
all exit relays.  One use case is HTTPS man-in-the-middle detection: A module
can fetch the certificate of a web server over all exit relays and then compare
its fingerprint with the expected, valid fingerprint.  Exposed attacks, however,
can be difficult to attribute because an attack can take place upstream of the
exit relay, e.g., at a malicious autonomous system.

In addition to the original modules that the exitmap authors shared with us, we
implemented exitmap modules to detect HTML tampering and TLS downgrading, by
connecting to servers under our control and raising an alert if the returned
HTML or TLS server hello were modified.  Our modules ran from August 2014 to
January 2016 and discovered 251 malicious exit relays, shown in
Appendix~\ref{sec:malicious-relays}, that we all reported to The Tor Project,
which subsequently blocked these relays.

\subsection{Threat model}
\label{sec:threat_model}
Most of this paper is on applying \sys to archived network data, but we can also
apply it to newly incoming data.  This puts us in an adversarial setting as
attackers can tune their Sybils to evade our system.  This is reflected in our
adversarial assumptions.  We assume that an adversary \emph{does} run more than
one Tor relay and exhibits redundancy in their relay configuration, or uptime
sequence.  An adversary further \emph{can} know how \sys's modules work, run
active or passive attacks, and make a limited effort to stay under the radar, by
diversifying parts of their configuration.  To detect Sybils, however, our
heuristics require \emph{some} redundancy.

\subsection{Analysis techniques}
\label{sec:techniques}
Having discussed our datasets and threat model, we now turn to presenting
techniques that can expose Sybils.  Our techniques are based on the insight that
Sybil relays typically \emph{behave or appear similarly}.  Shared configuration
parameters such as port numbers and nicknames cause similar appearance whereas
Sybils behave similarly when they reboot simultaneously, or exhibit identical
quirks when relaying traffic.

\begin{figure}[t]
	\centering
	\includegraphics[width=0.9\linewidth]{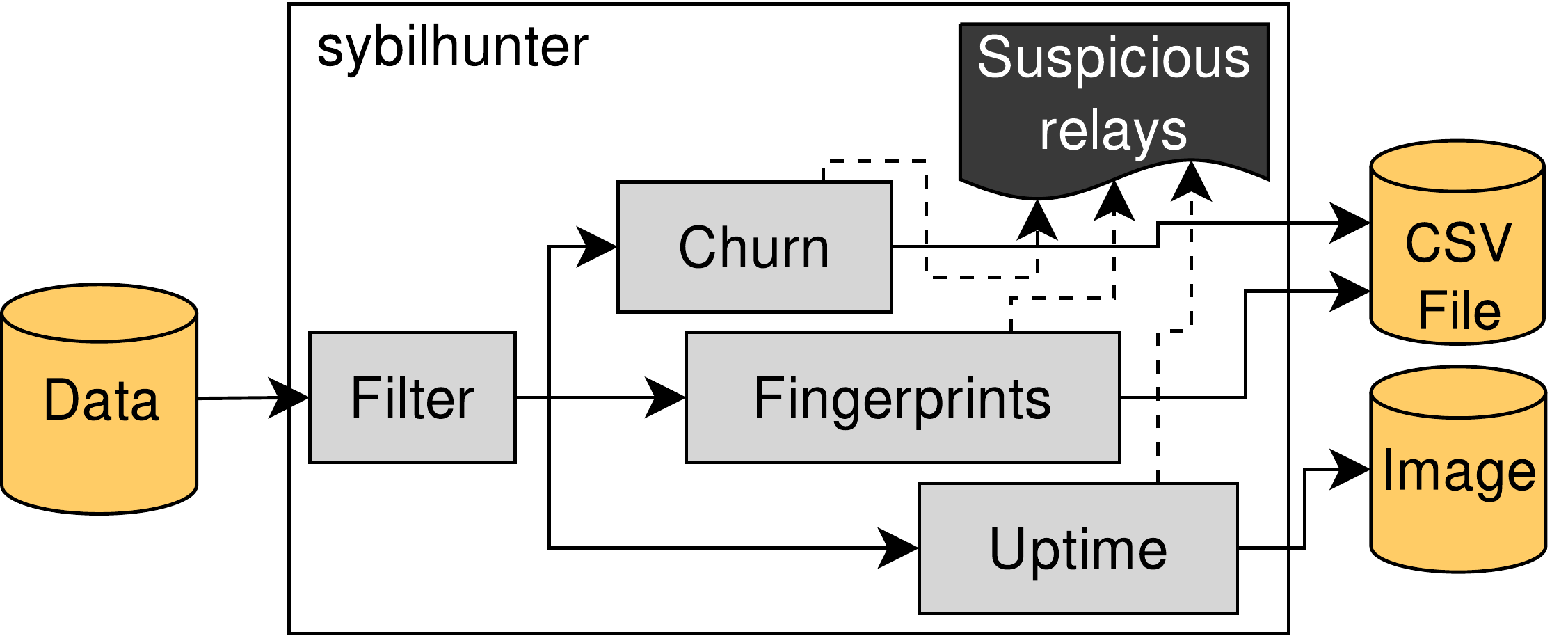}
	\caption{\Sys's internal architecture.}
	\label{fig:shr-internal}
\end{figure}

\Sys can analyze (\emph{i}) historical network data, dating back to 2007;
(\emph{ii}) online data, to detect new Sybils as they join the network; and
(\emph{iii}) find relays that might be associated with previously discovered,
malicious relays.  Figure~\ref{fig:shr-internal} shows \sys's internal
architecture.  Tor network data first passes a filtering component that can be
used to inspect a subset of the data.  It is then forwarded to one or more
modules that implement an analysis technique.  These modules work independently,
but share a data structure to find suspicious relays that show up in more than
one module.  Depending on the analysis technique, \sys's output is CSV
files or images.

While developing \sys, we had to make many design decisions that we
tackled by drawing on the experience we gained by manually analyzing numerous
Sybil groups.  We iteratively improved our code and augmented it with new
features when we experienced operational shortcomings.

\subsubsection{Network churn}
\label{sec:churn-time-series}
The churn rate of a distributed system captures the rate of joining and leaving
network participants.  In the Tor network, these participants are relays.  An
unexpectedly high churn rate between two subsequent consensuses means that many
relays joined or left, which can reveal Sybils and other network issues because
Sybil operators frequently start and stop their Sybils at the same time, to ease
administration---they behave similarly.

The Tor Project is maintaining a Python script~\cite{doctor} that determines the
number of previously unobserved relay fingerprints in new consensuses.  If that
number is greater than or equal to the static threshold 50, the script sends an
e-mail alert.  We reimplemented the script in \sys and ran it over all archived
consensus documents, dating back to 2007.  The script raised 47 alerts in nine
years, all of which seemed to be true positives, i.e., they should be of
interest to The Tor Project.  The script did not raise false positives,
presumably because the median number of new fingerprints in a consensus is only
six---significantly below the conservative threshold of 50.  Yet, the threshold
likely causes false negatives, but we cannot determine the false negative rate
because we lack ground truth.  In addition, The Tor Project's script does not
consider relays that left the network, does not distinguish between relays with
different flags, and does not adapt its threshold as the network grows.  We now
present an alternative approach that is more flexible and robust.

\begin{figure}[t]
	\centering
	\includegraphics[width=\linewidth]{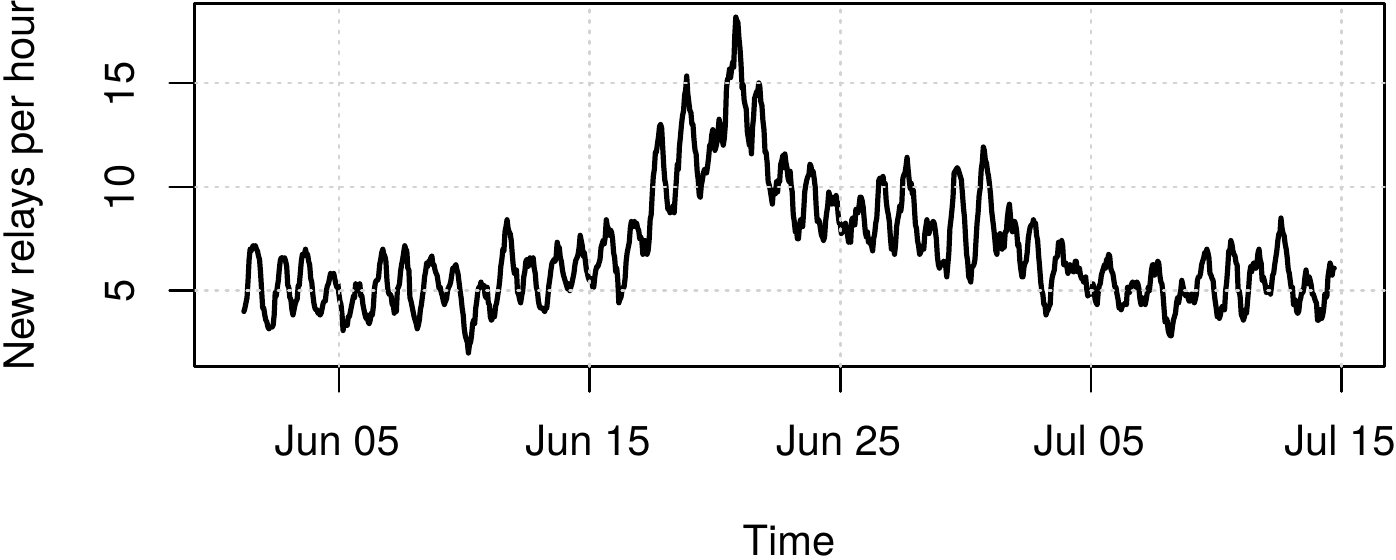}
	\caption{A flat hill of new relays in 2009.  The time series was smoothed
	using a moving average with a window size of 12 hours.}
	\label{fig:flat-hill}
\end{figure}

\begin{figure}[t]
	\centering
	\includegraphics[width=\linewidth]{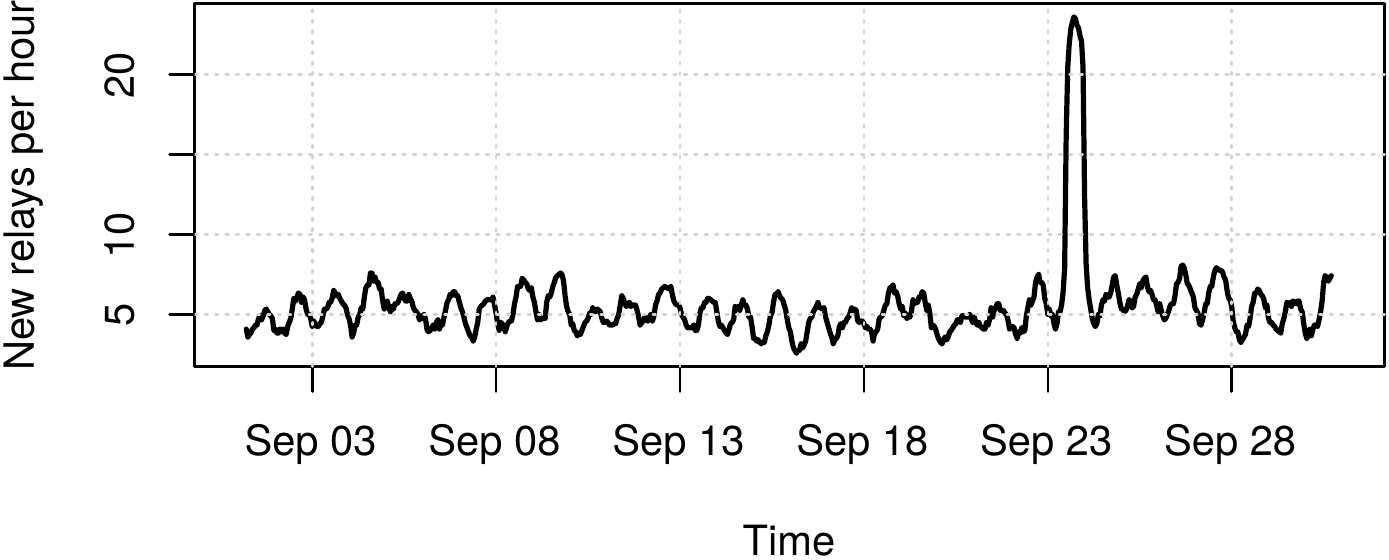}
	\caption{A sudden spike of new relays in 2010.  The time series was smoothed
	using a moving average with a window size of 12 hours.}
	\label{fig:sudden-spike}
\end{figure}

We found that churn anomalies worthy of our attention range from \emph{flat
hills} (Fig.~\ref{fig:flat-hill}) to \emph{sudden spikes}
(Fig.~\ref{fig:sudden-spike}).  Flat hills can be a sign of an event that
concerned a large number of relays, over many hours or days.  Such an event
happened shortly after the Heartbleed bug, when The Tor Project asked relay
operators to generate new keys.  Relay operators acted gradually, most within
two days.  Sudden spikes can happen if an attacker adds many relays, all at
once.  These are mere examples, however; the shape of a time series cannot tell
us anything about the nature of the underlying incident.

To quantify the churn rate $\alpha$ between two subsequent consensus documents,
we adapt Godfrey et al.'s formula, which yields a churn value that captures both
systems that joined and systems that left the network~\cite{Godfrey2006a}.
However, an unusually low number of systems that left could cancel out an
unusually high number of new systems and vice versa---an undesired property for
a technique that should spot abnormal changes.  To address this issue, we split
the formula in two parts, creating a time series for new relays ($\alpha_{n}$)
and for relays that left ($\alpha_{l}$).  $C_{t}$ is the network consensus at
time $t$, and $\setminus$ denotes the complement between two consensuses, i.e.,
the relays that are in the left operand, but not the right operand.  We define
$\alpha_{n}$ and $\alpha_{l}$ as

\begin{equation}
\alpha_{n} = \frac{\lvert C_{t} \setminus C_{t-1} \rvert}
{\lvert C_{t} \rvert}
\qquad\text{and}\qquad
\alpha_{l} = \frac{\lvert C_{t-1} \setminus C_{t} \rvert}
{\lvert C_{t-1} \rvert}.
\end{equation}

Both $\alpha_{n}$ and $\alpha_{l}$ are bounded to the interval $[0, 1]$.  A
churn value of 0 indicates no change between two subsequent consensuses whereas
a churn value of 1 indicates a complete turnover.  Determining $\alpha_{n,l}$
for the sequence $C_{t}, C_{t-1}$, \ldots, $C_{t-n}$, yields a time series of
churn values that can readily be inspected for abnormal spikes.  We found that
many churn anomalies are caused by relays that share a flag, or a flag
combination, e.g., \texttt{HSDir} (onion service directories) and \texttt{Exit}
(exit relays).  Therefore, \sys can also generate per-flag churn time series
that can uncover patterns that would be lost in a flag-agnostic time series.

Finally, to detect changes in the underlying time series trend---flat hills---we
can smooth $\alpha_{n,l}$ using a simple moving average $\lambda$ defined as

\begin{equation}
\lambda = \frac{1}{w} \cdot \sum_{i=0}^{w} \alpha_{i}.
\end{equation}

As we increase the window size $w$, we can detect more subtle changes in the
underlying churn trend.  If $\lambda$ or $\alpha_{n,l}$ exceed a manually
defined threshold, an alert is raised.  Section~\ref{sec:churn} elaborates on
how a threshold can be chosen in practice.

\subsubsection{Uptime matrix}
\label{sec:uptime-matrix}
For convenience, Sybil operators are likely to administer their relays
simultaneously, i.e., update, configure, and reboot them all at the same time.
This is reflected in their relays' uptime.  An operating system upgrade that
requires a reboot of Sybil relays will induce a set of relays to go offline and
return online in a synchronized manner.  To isolate such events, we are
visualizing the \emph{uptime patterns} of Tor relays by grouping together relays
whose uptime is highly correlated.  The churn technique presented above is
similar but it only provides an aggregate, high-level view on how Tor relays
join and leave the network.  Since the technique is aggregate, it is poorly
suited for visualizing the uptime of specific relays; an abnormally high churn
value attracts our attention but does not tell us what caused the anomaly.  To
fill this gap, we complement the churn analysis with an uptime matrix that we
will now present.

This uptime matrix consists of the uptime patterns of all Tor relays, which we
represent as binary sequences.  Each hour, when a new consensus is published, we
add a new data point---``online'' or ``offline''---to each Tor relay's sequence.
We visualize all sequences in a bitmap whose rows represent consensuses and
whose columns represent relays.  Each pixel denotes the uptime status of a
particular relay at a particular hour.  Black pixels mean that the relay was
online and white pixels mean that the relay was offline.  This type of
visualization was first proposed by Ensafi and subsequently implemented by
Fifield~\cite{Fifield2014a}.

\begin{figure}[t]
	\centering
	\includegraphics[width=\linewidth]{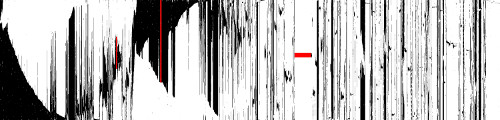}
	\caption{The uptime matrix for 3,000 Tor relays for all of November 2012.
	Rows represent consensuses and columns represent relays.  Black pixels mean
	that a relay was online, and white means offline.  Red blocks denote relays
	with identical uptime.}
	\label{fig:uptime-matrix}
\end{figure}

Of particular importance is how the uptime sequences are sorted.  If highly
correlated sequences are not adjacent in the visualization, we might miss them.
We sort sequences using single-linkage clustering, a type of hierarchical
clustering algorithm that forms groups bottom-up, based on the minimum distance
between group members.  Our clustering algorithm requires a distance function.
Similar to Andersen et al.~\cite{Andersen2002a}, we use Pearson's correlation
coefficient as our distance function because it tells us if two uptime sequences
change together.  The sample correlation coefficient $r$ yields a value in the
interval $[-1, 1]$.  A coefficient of $-1$ denotes perfect anti-correlation
(relay $R_1$ is only online when relay $R_2$ is offline) and 1 denotes perfect
correlation (relay $R_1$ is only online when relay $R_2$ is online).  We define
our distance function as $d(r) = 1 - r$, so two perfectly correlated sequences
have a distance of zero while two perfectly anti-correlated sequences have a
distance of two.  Once all sequences are sorted, we color adjacent sequences in
red if their uptime sequence is identical.  Figure~\ref{fig:uptime-matrix} shows
an example of our visualization algorithm, the uptime matrix for a subset of all
Tor relays in November 2012.

\subsubsection{Fingerprint analysis}
\label{sec:fingerprint-analysis}
The information a Tor client needs to connect to an onion service is stored in a
DHT that consists of a subset of all Tor relays, the onion service directories
(HSDirs).  As of February 2016, 46\% of all Tor relays serve as HSDirs.  A
daily-changing set of six HSDirs host the contact information of any given
onion service.  Tor clients contact one of these six HSDirs to request
information about the onion service they intend to connect to.  A HSDir becomes
responsible for an onion service if the difference between its relay fingerprint
and the service's descriptor ID is smaller than that of any other relay.  The
descriptor ID is derived from the onion service's public key, a time stamp, and
additional information.
All HSDirs are public, making it possible to determine at which position in the
DHT an onion service will end up at any point in the future.  Attackers can
exploit the ability to predict the DHT position by repeatedly generating
identity keys until their fingerprint is sufficiently close to the targeted
onion service's index, thus becoming its HSDir~\cite{Biryukov2013a}.

We detect relays that change their fingerprint frequently by maintaining a
lookup table that maps a relay's IP address to a list of all fingerprints we
have seen it use.  We sort the lookup table by the relays that changed their
fingerprints the most, and output the results.

\subsubsection{Nearest-neighbor search}
\label{sec:nearest-neighbor}
We frequently found ourselves in a situation where exitmap discovered a
malicious exit relay and we were left wondering if there were similar,
potentially associated relays.  Looking for such relays involved extensive
manual work, which we soon started to automate.  We needed an algorithm for
nearest-neighbor search that takes as input a ``seed'' relay and finds its $n$
most similar neighbors.  We define similarity as shared configuration parameters
such as port numbers, IP addresses, exit policies, or bandwidth values.  Our
search algorithm sorts relays by comparing these configuration parameters.

To quantify the similarity between two relays, we use the Levenshtein distance,
a distance metric that takes as input two strings and determines the minimum
number of modifications---insert, delete, and modify---that are necessary to
turn string $s_{2}$ into $s_{1}$.  Our algorithm turns the router statuses and
descriptors of two relays into strings and determines their Levenshtein
distance.  As an example, consider a simplified configuration representation
consisting of the concatenation of nickname, IP address, and port.  To turn
string $s_2$ into $s_1$, six operations are necessary; three modifications
(green) and two deletions (red):

\definecolor{change}{rgb}{0.7,1.0,0.7}
\definecolor{delete}{rgb}{1.0,0.7,0.7}

$s_1$: \texttt{Foo10.0.0.19001}

$s_2$: \texttt{\setlength{\fboxsep}{0pt}%
\colorbox{change}{\strut Bar}10.0.0.%
\colorbox{change}{\strut 2}%
\colorbox{delete}{\strut 54}%
9001}

Our algorithm determines the Levenshtein distance between a ``seed'' relay and
all $n-1$ relays in a consensus.  It then sorts the calculated distances in
ascending order and prints the most similar relays to the console.  For a
consensus consisting of 6,525 relays, our algorithm takes approximately 1.5
seconds to finish.\footnote{We measured on an Intel Core i7-3520M CPU at 2.9
GHz, a consumer-grade CPU.}

\section{Evaluation and results}
\label{sec:results}
Equipped with \sys, we applied our techniques to nine years of archived Tor
network data, resulting in several megabytes of CSV files and uptime images.  We
sorted our results in descending order by severity, and started manually
analyzing the most significant incidents.  Several outliers were caused by
problems and events in the Tor network that were unrelated to Sybil relays.
Instead of providing an exhaustive list of all potential Sybils, we focus on our
most salient findings---relay groups that were either clearly malicious or
distinguished themselves otherwise.\footnote{Our datasets and visualizations are
available online, and can be inspected for an exhaustive list of potential
Sybils.  The URL is \url{https://nymity.ch/sybilhunting/}.}

Once we discovered a seemingly harmful Sybil group, we reported it to The Tor
Project.  To defend against Sybil attacks, directory authorities can either
remove a relay from the consensus, or take away its \texttt{Valid} flag, which
means that the relay is still in the consensus, but Tor clients will not
consider it for their first or last hop in a circuit.  The majority of directory
authorities, i.e., five out of eight, must agree on either strategy.  This
mechanism is meant to distribute the power of removing relays into the hands of
a diverse set of people.

We present our results by first giving an overview of the most interesting
Sybils we discovered in Section~\ref{sec:sybil_groups}; followed by
technique-specific results in Sections~\ref{sec:churn}, \ref{sec:uptime}, and
\ref{sec:fingerprints}; an evaluation of our nearest-neighbor search in
Section~\ref{sec:accuracy}; and the computational cost of our techniques in
Section~\ref{sec:performance}.

\subsection{Sybil characterization}
\label{sec:sybil_groups}
Table~\ref{tab:sybils} shows the most interesting Sybil groups we identified.
The columns show (\emph{i}) what we believe to be the purpose of the Sybils,
(\emph{ii}) when the Sybil group was at its peak size, (\emph{iii}) the ID we
gave the Sybils, (\emph{iv}), the number of Sybil fingerprints, (\emph{v}) the
analysis techniques that could discover the Sybils, and (\emph{vi}) a short
description.  The analysis techniques are abbreviated as ``E'' (exitmap), ``C''
(Churn), ``U'' (Uptime), ``F'' (Fingerprint), and ``N'' (Neighbor search).  We
now discuss the most insightful incidents in greater detail.

\begin{table*}[ht!]
\small
\centering
\begin{tabularx}{\textwidth}{c|c c c c X}
\hline
\textbf{Purpose} & \textbf{Peak activity} & \textbf{Group ID} & \textbf{Number} &
\textbf{Method} & \textbf{Description} \\
\hline
\multirow{6}{*}{\textbf{MitM}}
& Jan 2016 & rewrite$\dagger$ & 42 & E & Replaced onion domains with
impersonation site. \\

& Nov 2015 & rewrite$\dagger$ & 8 & E & Replaced onion domains with impersonation
site. \\

& Jun 2015 & rewrite$\dagger$ & 55 & E & Replaced onion domains with
impersonation site. \\

& Apr 2015 & rewrite$\dagger$ & 71 & U,E & Replaced onion domains with
impersonation site. \\

& Mar 2015 & redirect$\ddagger$ & 24 & E & Redirected users to impersonated site.
\\

& Feb 2015 & redirect$\ddagger$ & 17 & E & Redirected users to impersonated site.
\\

& Jan 2015 & redirect$\ddagger$ & 26 & E & Redirected users to impersonated site.
\\

\hline

\multirow{2}{*}{\textbf{Botnet}}
& Mar 2014 & default & --- & N & Likely a Windows-powered botnet.  The group
features wide geographical distribution, which is uncommon for typical Tor
relays. \\

& Oct 2010 & trotsky & 649 & N & The relays were likely part of a botnet.  They
appeared gradually, and were all running Windows. \\

\hline

\multirow{5}{*}{\textbf{Unknown}}
& Jan 2016 & cloudvps & 61 & C,U & Hosted by Dutch hoster XL Internet Services. \\

& Nov 2015 & 11BX1371 & 150 & C,U & All relays were in two /24 networks and a single
relay had the \texttt{Exit} flag.  \\

& Jul 2015 & DenkoNet & 58 & U & Hosted on Amazon AWS and only present in a single
consensus.  No relay had the \texttt{Exit} flag. \\

& Jul 2015 & cloudvps & 55 & C,U & All relays only had the \texttt{Running} and
\texttt{Valid} flag.  As their name suggests, the relays were hosted by the
Dutch hoster ``CloudVPS.'' \\

& Dec 2014 & Anonpoke & 284 & C,U & The relays did not have the \texttt{Exit} flag
and were removed from the network before they could get the \texttt{HSDir} flag.
\\

& Dec 2014 & FuslVZTOR & 246 & C,U & The relays showed up only hours after the
LizardNSA incident. \\

\hline

\multirow{1}{*}{\textbf{DoS}}
& Dec 2014 & LizardNSA & 4,615 & C,U & A group publicly claimed to be
responsible for the attack~\cite{lizards}.  All relays were hosted in the Google
cloud and The Tor Project removed them within hours. \\

\hline

\multirow{4}{*}{\textbf{Research}}
& May 2015 & fingerprints & 168 & F & All twelve IP addresses, located in the
same /24, changed their fingerprint regularly, presumably in an attempt to
manipulate the distributed hash table. \\

& Mar 2014 & FDCservers & 264 & C,U & Relays that were involved in an
experimental onion service deanonymization attack~\cite{cmucert}. \\

& Feb 2013 & AmazonEC2 & 1,424 & F,C,U & We observed 1,424 relay fingerprints on
88 IP addresses.  These Sybils were likely part of a research
project~\cite{Biryukov2013a}. \\

& Jun 2010 & planetlab & 595 & C,U & According to a report from The Tor
Project~\cite{progressreport}, a researcher started these relays to learn more
about scalability effects. \\

\hline

\end{tabularx}
\caption{The Sybil groups we discovered using \sys and our exitmap
modules.  We believe that groups marked with the
symbols $\dagger$ and $\ddagger$ were run by the same operator, respectively.}
\label{tab:sybils}
\end{table*}

\paragraph{The ``rewrite'' Sybils}
These recurring Sybils hijacked Bitcoin transactions by rewriting Bitcoin
addresses.  All relays had the \texttt{Exit} flag and replaced onion domains
found in a web server's HTTP response with an impersonation domain, presumably
hosted by the attacker.  Interestingly, the impersonation domains shared a
prefix with the original.  For example, \textbf{sigaint}evyh2rzvw.onion was
replaced with the impersonation domain \textbf{sigaint}z7qjj3val.onion whose
first seven digits are identical to the original.  The attacker could create
shared prefixes by repeatedly generating key pairs until the hash over the
public key resembled the desired prefix.  Onion domains are generated by
determining the SHA-1 hash over the public key, truncating it to its 80 most
significant bits, and encoding it in Base32.  Each Base32 digit of the
16-digit-domain represents five bits.  Therefore, to get an $n$-digit prefix in
the onion domain, $2^{5 n - 1}$ operations are required on average.  For the
seven-digit prefix above, this results in $2^{5 \cdot 7 - 1} = 2^{34}$
operations.  The author of scallion~\cite{scallion}, a tool for generating
vanity onion domains, determined that an nVidia Quadro K2000M, a mid-range
laptop GPU, is able to generate 90 million hashes per second.  On this GPU, a
partial collision for a seven-digit prefix can be found in $2^{34} \cdot
\frac{1}{90,000,000} \simeq 190$ seconds, i.e., just over three minutes.

We inspected some of the phishing domains and found that the attackers replaced
the original Bitcoin addresses, presumably with addresses under their control,
to hijack transactions.  Therefore, we believe that this attack was financially
motivated.

\paragraph{The ``redirect'' Sybils}
These relays all had the \texttt{Exit} flag and tampered with HTTP redirects of
exit traffic.  Some Bitcoin sites would redirect users from their HTTP site to
the encrypted HTTPS version, to protect their users' login credentials.  This
Sybil group tampered with the redirect and directed users to an impersonation
site, resembling the original Bitcoin site, perhaps to steal credentials.  We
only observed this attack for Bitcoin sites, but cannot rule out that other
sites were not attacked.

Interestingly, the Sybils' descriptors and consensus entries had less in common
than other Sybil groups.  They used a small set of different ports, Tor
versions, bandwidth values, and their nicknames did not exhibit an
easily-recognizable pattern.  In fact, the only reason why we know that these
Sybils belong together is because their attack was identical.

We discovered three Sybil groups that implemented the redirect attack, each of
them beginning to surface when the previous one got blocked.  The initial group
first showed up in May 2014, with only two relays, but slowly grew over time,
until it was finally discovered in January 2015.  We believe that these Sybils
were run by the same attacker because their attack was identical.

It is possible that this Sybil group was run by the same attackers that
controlled the ``rewrite'' group but we have no evidence to support that
hypothesis.  Interestingly, only our exitmap module was able to spot these
Sybils.  The relays joined the network gradually over time and had little in
common in their configuration, which is why our \sys methods failed.  In fact,
we cannot rule out that the adversary was upstream, or gained control over these
relays.

\paragraph{The ``FDCservers'' Sybils}
These Sybils were used to deanonymize onion service users, as discussed by The
Tor Project in a July 2014 blog post~\cite{cmucert}.  Supposedly, CMU/CERT
researchers were executing a traffic confirmation attack by sending sequences of
\texttt{RELAY\_EARLY} and \texttt{RELAY} cells as a signal down the circuit to
the client, which the reference implementation never
does~\cite{cmucert,cmucert2}.  The attacking relays were onion service
directories and guards, allowing them to control both ends of the circuit for
some Tor clients that were fetching onion service descriptors.  Most relays were
running FreeBSD, used Tor in version 0.2.4.18-rc, had identical flags, mostly
identical bandwidth values, and were located in 50.7.0.0/16 and 204.45.0.0/16.
All of these shared configuration options made the relays easy to identify.

The relays were added to the network in batches, presumably starting in October
2013.  On January 30, 2014, the attackers added 58 relays to the 63 existing
ones, giving them control over 121 relays.  On July 8, 2014, The Tor Project
blocked all 123 IP addresses that were running at the time.

\paragraph{The ``default'' Sybils}
This group, named after the Sybils' shared nickname ``default,'' has been around
since September 2011 and consists of Windows-powered relays only.  We extracted
relays by filtering consensuses for the nickname ``default,'' onion routing port
443, and directory port 9030.  The group features high IP address churn.  For
October 2015, we found ``default'' relays in 73 countries, with the top three
countries being Germany~(50\%), Russia~(8\%), and Austria~(7\%).  The majority
of these relays, however, has little uptime.
Figure~\ref{fig:default-sybils-uptime} shows the uptime matrix for ``default''
relays in October 2015.  Many relays exhibit a diurnal pattern, suggesting that
they are powered off regularly---as it often is the case for desktop computers
and laptops.

\begin{figure}[t]
	\centering
	\includegraphics[width=\linewidth]{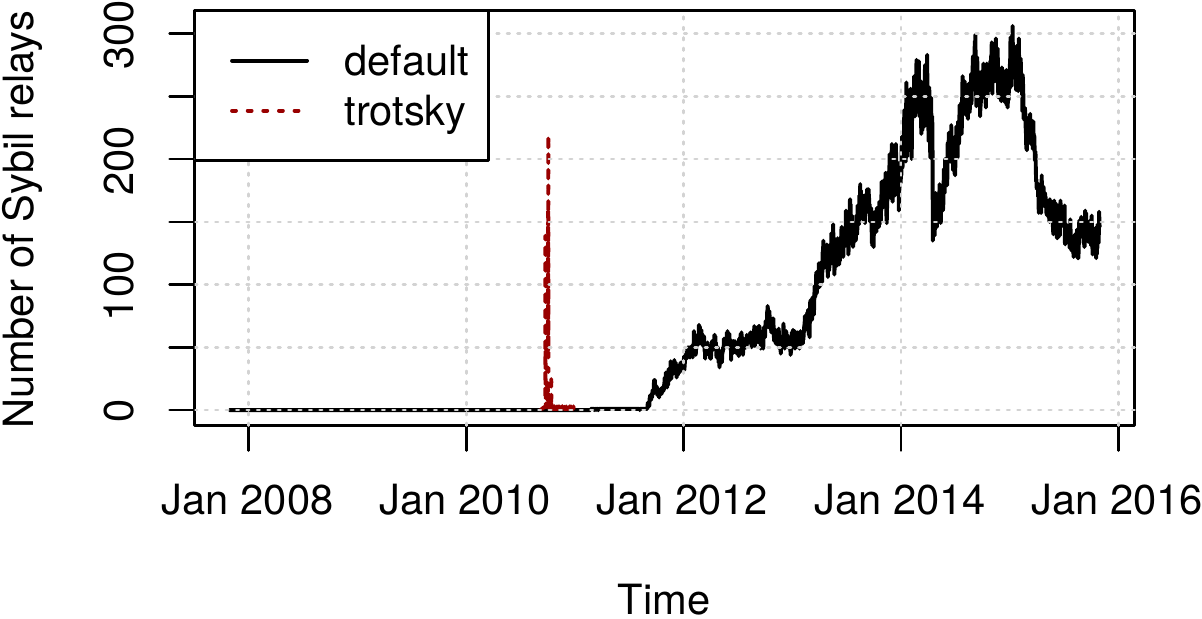}
	\caption{The number of ``default'' and ``trotsky'' Sybil members over time.}
	\label{fig:default-over-time}
\end{figure}

To get a better understanding of the number of ``default'' relays over time, we
analyzed all consensuses, extracting the number of relays whose nickname was
``default,'' whose onion routing port was 443, and whose directory port was
9001.  We did this for the first consensus every day and plot the result in
Figure~\ref{fig:default-over-time}.  Note that we might overestimate the numbers
as our filter could capture unrelated relays.

The above suggests that some of the ``default'' relays are running without the
owner's knowledge.  While the relays do not fit the pattern of Sefnit (a.k.a.
Mevade)~\cite{sefnit} and Skynet~\cite{skynet}---two pieces of malware that use
an onion service as command and control server---we believe that the ``default''
relays constitute a botnet.

\paragraph{The ``trotsky'' Sybils}
Similar to the ``default'' group, the ``trotsky'' relays appear to be part of
a botnet.  Most of the relays' IP addresses were located in Eastern Europe, in
particular in Slovenia, Croatia, and Bosnia and Herzegovina.  The relays were
all running on Windows, in version 0.2.1.26, and listening on port 443.  Most of
the relays were configured as exits, and The Tor Project assigned some of them
the \texttt{BadExit} flag.

The first ``trotsky'' members appeared in September 2010.  Over time, there were
two relay peaks, reaching 139 (September 23) and 219 (October 3) relays, as
illustrated in Figure~\ref{fig:default-over-time}.  After that, only 1--3 relays
remained in the consensus.

\paragraph{The ``Amazon EC2'' Sybils}
The relays all used randomly-generated nicknames, consisting of sixteen or
seventeen letters and numbers; Tor in version 0.2.2.37; GNU/Linux; and IP
addresses in Amazon's EC2 netblock.  Each of the 88 IP addresses changed its
fingerprint 24 times, but not randomly: the fingerprints were chosen
systematically, in a small range.  For example, relay 54.242.248.129 had
fingerprints with the prefixes \texttt{8D}, \texttt{8E}, \texttt{8F}, and
\texttt{90}.  The relays were online for 48 hours.  After 24 hours, most of the
relays obtained the \texttt{HSDir} flag.

We believe that this Sybil group was run by Biryukov, Pustogarov, and Weinmann
as part of their Security and Privacy 2013 paper ``Trawling for Tor Hidden
Services''~\cite{Biryukov2013a}---one of the few Sybil groups that were likely
run by academic researchers.

\paragraph{The ``FuslVZTOR'' Sybils}
All machines were middle relays and hosted in 212.38.181.0/24, which is owned by
a UK VPS provider.  The directory authorities started rejecting the relays five
hours after they joined the network.  The relays advertized the default bandwidth
of 1 GiB/s and used randomly determined ports.  The Sybils were active in
parallel to the ``LizardNSA'' attack, but there is no reason to believe that
both incidents were related.

\paragraph{The ``Anonpoke'' Sybils}
All relays shared the nickname ``Anonpoke'' and were online for four hours until
they were rejected.  All relays were hosted by a VPS provider in the U.S.,
Rackspace, with the curious exception of a single relay that was hosted in the
UK, and running a different Tor version.  The relays advertized the default
bandwidth of 1 GiB/s on port 9001 and 9030.  All relays were middle relays and
running as directory mirror.  All Sybils were configured to be an onion service
directory, but did not manage to get the flag in time.

\paragraph{The ``PlanetLab'' Sybils}
A set of relays that used a variation of the strings ``planet'', ``plab'',
``pl'', and ``planetlab'' as their nickname.  The relays' exit policy allowed
ports 6660--6667, but they did not get the \texttt{Exit} flag.  The Sybils were
online for three days and then removed by The Tor Project, as mentioned in a
blog post~\cite{progressreport}.  The blog post further says that the relays
were run by a researcher to learn more about ``cloud computing and scaling
effects.''

\paragraph{The ``LizardNSA'' Sybils}
All relays were hosted in the Google Cloud and only online for ten hours, until
the directory authorities started to reject them.  The majority of machines were
middle relays (96\%), but the attackers also started some exit relays (4\%).
The Sybils were set up to be onion service directories, but the relays were
taken offline before they could earn the \texttt{HSDir} flag.  If all relays
would have obtained the \texttt{HSDir} flag in time, they would have constituted
almost 50\% of all onion service directories; the median number of onion
service directories on December 26 was 3,551.

Shortly after the attack began, somebody claimed responsibility on the tor-talk
mailing list~\cite{lizards}.  Judging by the supposed attacker's demeanor, the
attack was mere mischief.

\subsection{Churn rate analysis}
\label{sec:churn}
\begin{figure}[t]
	\centering
	\includegraphics[width=\linewidth]{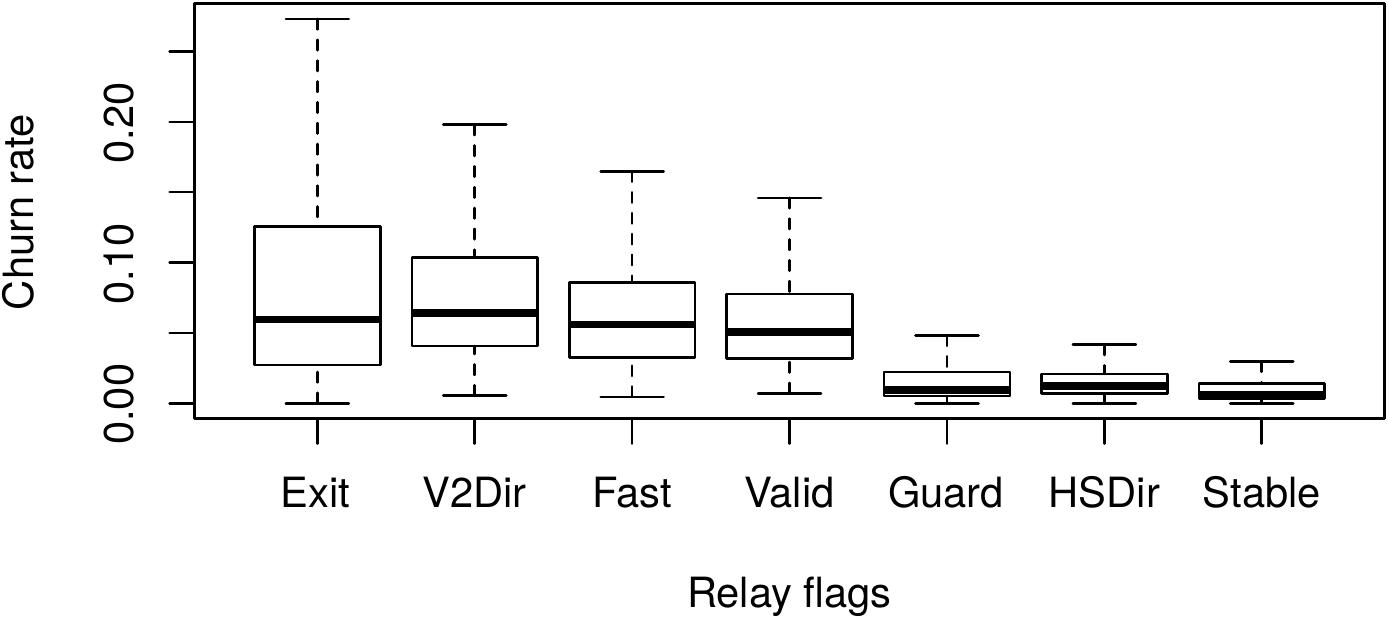}
	\caption{The churn distribution for seven relay flags.  We removed values
		greater than the plot whiskers.}
	\label{fig:churn-boxplot}
\end{figure}

We determined the churn rate between two subsequent consensuses for all 72,061
consensuses that were published between October 2007 and January 2016.
Considering that (\emph{i}) there are 162 gaps in the archived data, that
(\emph{ii}) we create time series for joining and leaving relays, and that
(\emph{iii}) we determined churn values for all twelve relay flags, we ended up
with $(72,061 - 162) \cdot 2 \cdot 12 = 1,725,576$ churn values.
Figure~\ref{fig:churn-boxplot} shows a box plot for the churn distribution
(joining and leaving churn values concatenated) for the seven most relevant
relay flags.  We removed values greater than the plot whiskers (which extend to
values $1.5$ times the interquartile range from the box) to better visualize the
width of the distributions.  Unsurprisingly, relays with the \texttt{Guard},
\texttt{HSDir}, and \texttt{Stable} flag experience the least churn, probably
because relays are only awarded these flags if they are particularly stable.
Exit relays have the most churn, which is surprising given that exit relays are
particularly sensitive to operate.  Interestingly, the median churn rate of the
network has steadily decreased over the years, from $0.04$ in 2008 to $0.02$ in
2015.

\begin{figure}[t]
	\centering
	\includegraphics[width=\linewidth]{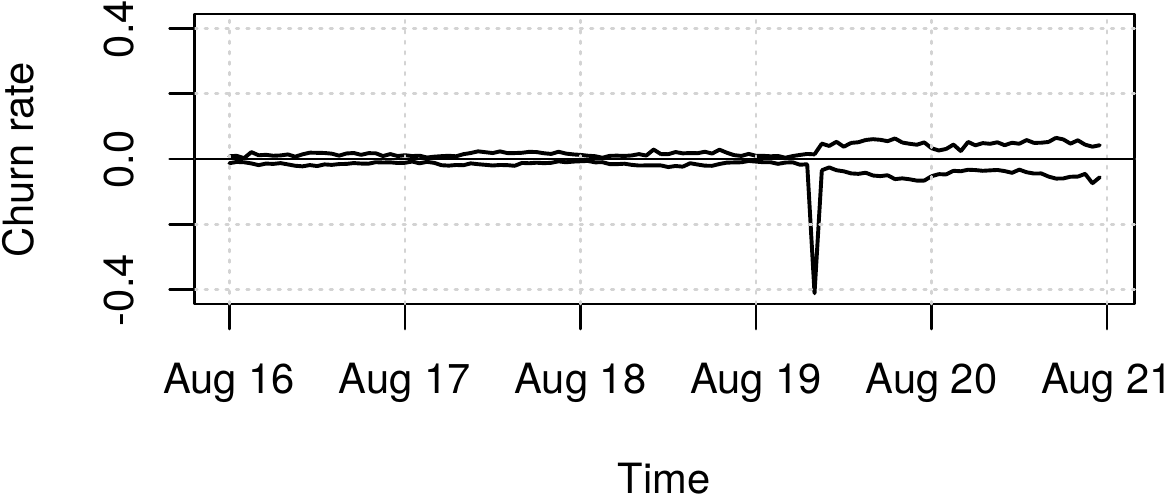}
	\caption{In August 2008, an upgrade in Tor's consensus format caused the
	biggest anomaly in our dataset.  The positive time series represents relays
	that joined and the negative one represents relays that left.}
	\label{fig:2008-08}
\end{figure}

Figure~\ref{fig:2008-08} illustrates churn rates for five days in August 2008,
featuring the most significant anomaly in our data.  On August 19, 822 relays
left the network, resulting in a sudden spike, and an increase in the baseline.
The spike was caused by the Tor network's switch from consensus format version
three to four.  The changelog says that in version four, routers that
do not have the \texttt{Running} flag are no longer listed in the consensus.

\begin{figure}[t]
	\centering
	\includegraphics[width=0.9\linewidth]{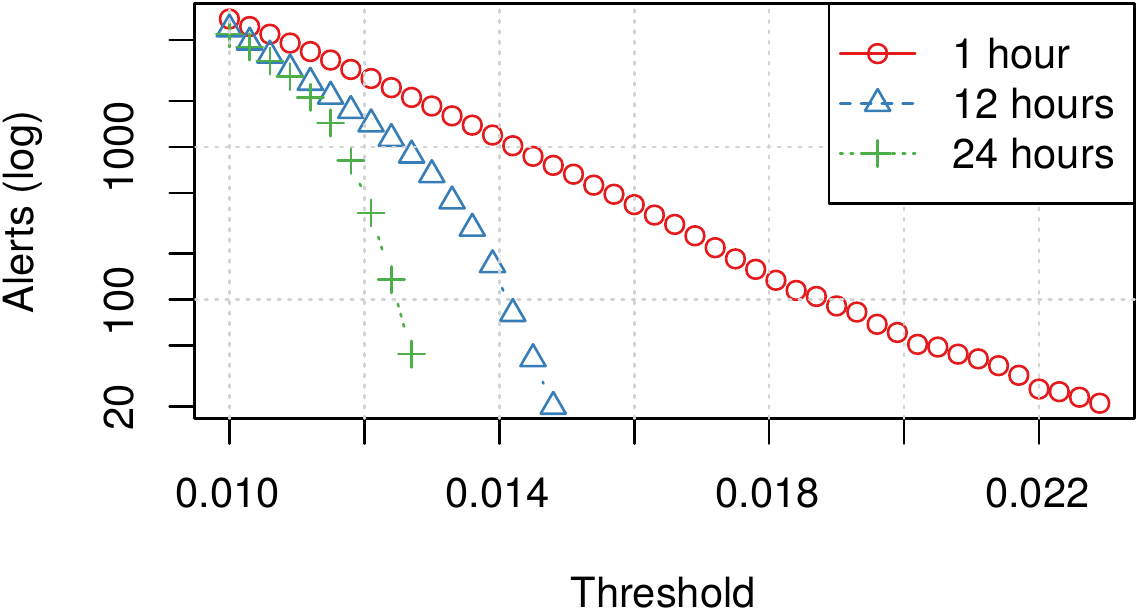}
	\caption{The number of alerts (in log scale) in 2015 as the detection
	threshold increases, for three smoothing window sizes.}
	\label{fig:threshold-alarm}
\end{figure}

To alleviate the choice of a detection threshold, we plot the number of alerts
(in log scale) in 2015 as the threshold increases.  We calculate these numbers
for four simple moving average window sizes.  The result is shown in
Figure~\ref{fig:threshold-alarm}.  Depending on the window size, thresholds
greater than 0.012 seem practical considering that 181 alerts per year average
to approximately one alert in two days---a tolerable number of incidents to
investigate.  Unfortunately, we are unable to determine the false positive rate
because we do not have ground truth.

\subsection{Uptime analysis}
\label{sec:uptime}
We generated relay uptime visualizations for each month since 2007, resulting in
100 images.  We now discuss a subset of these images that contain particularly
interesting patterns.

\begin{figure}[t]
	\centering
	\includegraphics[width=\linewidth]{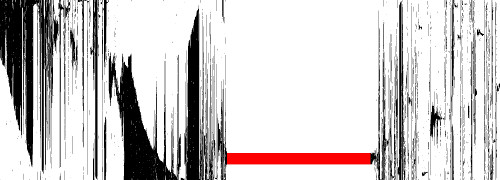}
	\caption{In June 2010, a researcher started several hundred Tor relays on
		PlanetLab~\cite{progressreport}.  Die image shows the uptime of 2,000
		relays for all of June.}
	\label{fig:2010-06-planetlab}
\end{figure}

Figure~\ref{fig:2010-06-planetlab} shows June 2010, featuring a clear ``Sybil
block'' on the left side.  The Sybils belonged to a researcher who, as
documented by The Tor Project~\cite{progressreport}, started several hundred Tor
relays on PlanetLab for research on scalability.  Our manual analysis could
verify this.  The relays were easy to identify because their nicknames suggested
that they were hosted on PlanetLab, containing strings such as ``planetlab,''
``planet,'' and ``plab.''  Note the small height of the Sybil block, indicating
that the relays were only online for a short time.

\begin{figure}[t]
	\centering
	\includegraphics[width=\linewidth]{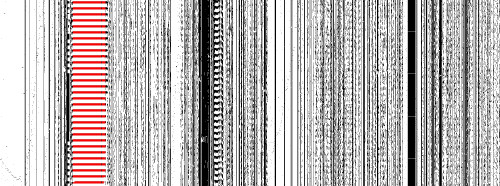}
	\caption{August 2012 featured a curious ``step pattern,'' caused by
	approximately 100 Sybils.  The image shows the uptime of 2,000 relays for
	all of August.}
	\label{fig:2012-08-steppattern}
\end{figure}

Figure~\ref{fig:2012-08-steppattern} features a curious ``step pattern'' for
approximately 100 relays, all of which were located in Russia and Germany.  The
relays appeared in December 2011, and started exhibiting the diurnal step
pattern (nine hours uptime followed by fifteen hours downtime) in March 2012.
All relays had similar nicknames, consisting of eight seemingly
randomly-generated characters.  In April 2013, the relays finally disappeared.

\begin{figure}[t]
	\centering
	\includegraphics[width=\linewidth]{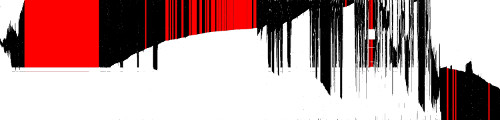}
	\caption{In April 2014, the Heartbleed bug forced The Tor Project to reject
	many affected relays.  The image shows the uptime of 3,000 relays for all of
	April.}
	\label{fig:2014-04-heartbleed}
\end{figure}

Figure~\ref{fig:2014-04-heartbleed} shows the effect of the Heartbleed
incident~\cite{Durumeric2014a} on the Tor network.  Several days after the
incident, The Tor Project decided to block all relays that had not generated new
key pairs.  The large red rectangle on the left side of the image illustrates
when the biggest part of the block became active, rejecting approximately 1,700
Tor relay fingerprints.

\begin{figure}[t]
	\centering
	\includegraphics[width=\linewidth]{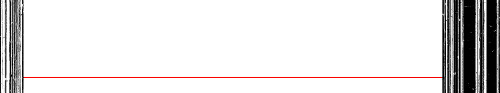}
	\caption{In December 2014, a group of attacker started several hundred Tor
		relays in the Google cloud.  The image shows the uptime of 4,000 relays
		for all of December.}
	\label{fig:2014-12-lizard}
\end{figure}

Figure~\ref{fig:2014-12-lizard} illustrates the largest Sybil group to date,
comprising 4,615 Tor relays that an attacker started in the Google cloud in
December 2014.  Because of its magnitude, the attack was spotted almost
instantly, and The Tor Project removed the offending relays only ten hours
after they appeared.

\subsection{Fingerprint anomalies}
\label{sec:fingerprints}
\begin{figure}[t]
	\centering
	\includegraphics[width=0.9\linewidth]{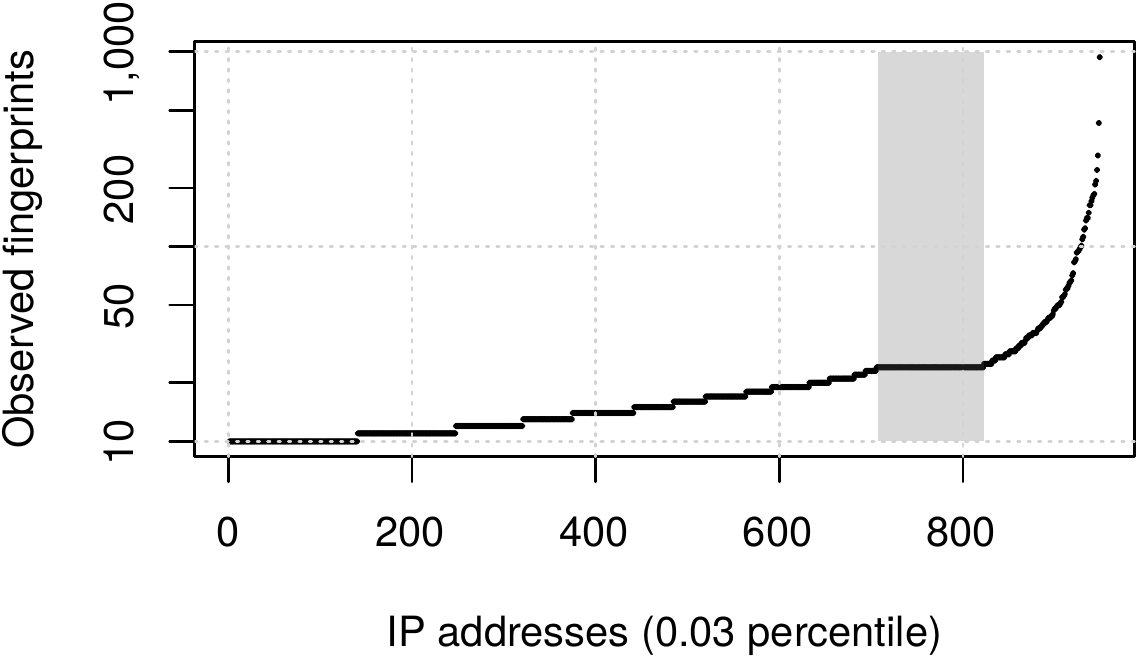}
	\caption{The number of observed fingerprints for the 1,000 relays that
	changed their fingerprints the most.}
	\label{fig:fingerprints}
\end{figure}

We determined how often all Tor relays changed their fingerprint from 2007 to
2015.  Figure~\ref{fig:fingerprints} illustrates the number of fingerprints ($y$
axis) we have observed for the 1,000 Tor relays ($x$ axis) that changed their
fingerprint the most.  All these relays changed their fingerprint at least ten
times.  Twenty one relays changed their fingerprint more than 100 times, and the
relay at the very right end of the distribution changed its fingerprint 936
times.  This relay's nickname was ``openwrt,'' suggesting that it was a home
router that was rebooted regularly.  It was running from August 2010 to December
2010.

Figure~\ref{fig:fingerprints} further contains a peculiar plateau, shown in the
shaded area between index 707 and 803.  This plateau was caused by a group of
Sybils, hosted in Amazon EC2, that changed their fingerprint exactly 24 times.
Upon inspection, we noticed that this was likely an experiment for a Security
and Privacy 2013 paper on deanonymizing Tor onion services~\cite{Biryukov2013a}.

We also found that many IP addresses in the range 199.254.238.0/24 changed their
fingerprint frequently.  We contacted the owner of the address block and were
told that the block used to host VPN services.  Apparently, several people
started Tor relays and since the VPN service would not assign permanent IP
addresses, the Tor relays would periodically change their address, causing the
churn we observe.

\subsection{Accuracy of nearest-neighbor search}
\label{sec:accuracy}
Given a single Sybil relay, how good is our nearest-neighbor search at finding
the remaining Sybils?  To answer this question, we now evaluate our algorithm's
\emph{accuracy}, which we define as the fraction of neighbors it correctly
labels as Sybils.  For example, if eight out of ten Sybils are correctly labeled
as neighbors, the accuracy is 0.8.

A sound evaluation requires ground truth, i.e., relays that are \emph{known} to
be Sybils.  All we have, however, are relays that we \emph{believe} to be
Sybils.  In addition, the number of Sybils we found is only a lower bound---we
are unlikely to have detected all Sybil groups.  Therefore, our evaluation is
doomed to overestimate our algorithm's accuracy because we are unable to test it
on the Sybils we did not discover.

We evaluate our search algorithm on two datasets; the ``bad exit'' Sybil groups
from Table~\ref{tab:exitmap-dataset}, and relay families.  We chose the bad exit
Sybils because we observed them running identical, active attacks, which makes
us confident that they are in fact Sybils.  Recall that a relay family is a set
of Tor relays that is controlled by a single operator, but configured to express
this mutual relationship in the family members' configuration file.  Relay
families are benign Sybils.  As of January 2016, approximately 400 families
populate the Tor network, ranging in size from only two to 25 relays.

We evaluate our algorithm by finding the nearest neighbors of a family member.
Ideally, all neighbors are family members, but the use of relay families as
ground truth is very likely to overestimate results because family operators
frequently configure their relays identically on purpose.  At the time of this
writing, a popular relay family has the nicknames ``AccessNow000'' to
``AccessNow009,'' adjacent IP addresses, and identical contact
information---perfect prerequisites for our algorithm.  We expect the operators
of malicious Sybils, however, to go out of their way to obscure the relationship
between their relays.

\begin{figure}
\centering
\subfigure[Bad exit relay Sybils]{
	\includegraphics[width=0.46\linewidth]{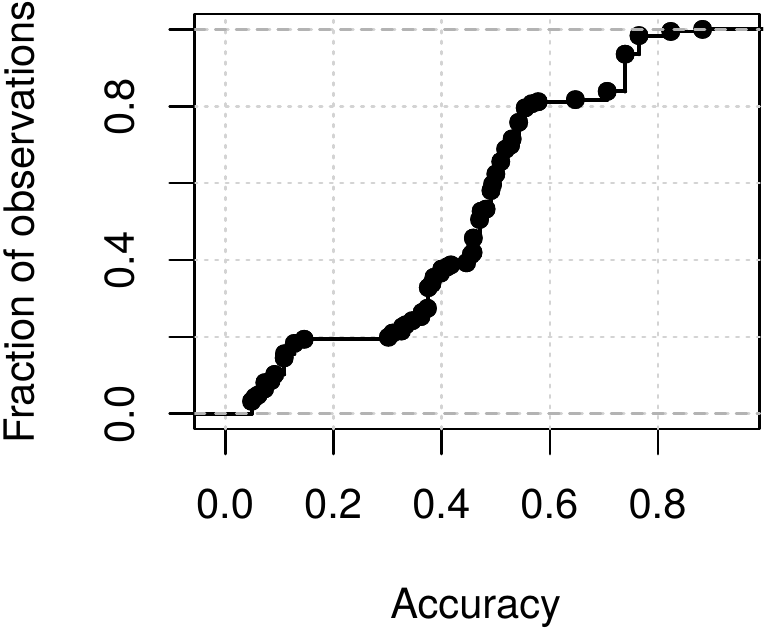}
\label{fig:badexit-accuracy}
}
\subfigure[Benign family Sybils]{
	\includegraphics[width=0.46\linewidth]{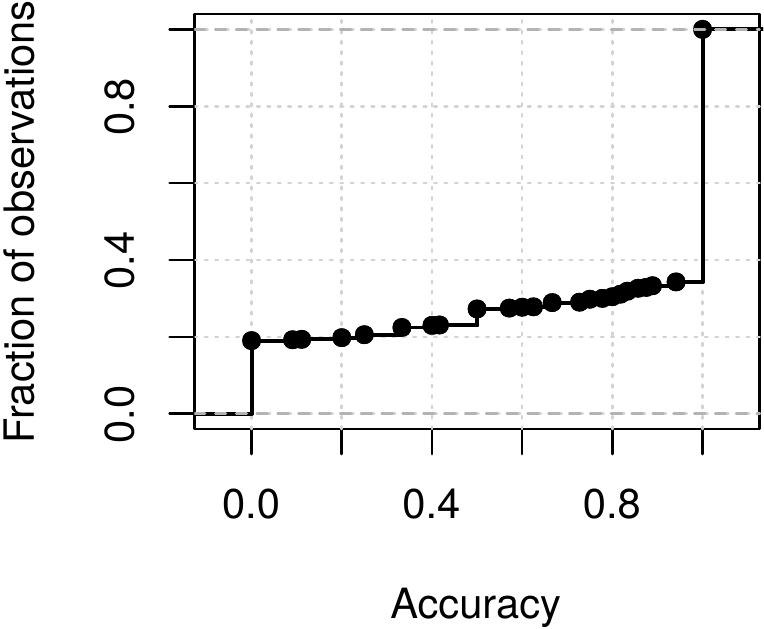}
\label{fig:family-accuracy}
}
\caption{ECDF for our two evaluations, the bad exit Sybils
	in Fig.~\ref{fig:badexit-accuracy} and the benign family Sybils
	in Fig.~\ref{fig:family-accuracy}.}
\label{fig:accuracy}
\end{figure}

To determine our algorithm's accuracy, we used all relay families that were
present in the first consensus that was published in October 2015.  For each
relay that had at least one mutual family relationship, we searched for its $n -
1$ nearest neighbors where $n$ is the family size.  Basically, we evaluated how
good our algorithm is at finding the relatives of a family member.  We
determined the accuracy---a value in $[0,1]$---for each family member.  The
result is shown in Figure~\ref{fig:family-accuracy}, a distribution of accuracy
values.

Next, we repeated the evaluation with the bad exit Sybil groups from
Table~\ref{tab:exitmap-dataset}.  Again, we searched for the $n - 1$ nearest
neighbors of all bad exit relays, where $n$ is the size of the Sybil group.  The
accuracy is the fraction of relays that our algorithm correctly classified as
neighbor.  The result is illustrated in Figure~\ref{fig:badexit-accuracy}.

As expected, our algorithm is significantly more accurate for the family
dataset---66\% of searches had perfect accuracy.  The bad exit dataset, however,
did worse.  Not a single search had perfect accuracy and 59\% of all searches
had an accuracy in the interval $[0.3,0.6]$.  Nevertheless, we find that our
search algorithm facilitates manual analysis given how quickly it can provide us
with a list of the most similar relays.  Besides, false positives (i.e.,
neighbors that are not Sybils) are cheap as \sys users would not spend much time
on neighbors that bear little resemblance to the ``seed'' relay.

\subsection{Computational cost}
\label{sec:performance}
Fast techniques lend themselves to being run hourly, for every new consensus,
while slower ones must be run less frequent.  Table~\ref{tab:exp-deployment}
gives an overview of the runtime of our methods.\footnote{We determined all
performance numbers on an Intel Core i7-3520M CPU at 2.9 GHz, a consumer-grade
CPU.}  We stored our datasets on a solid state drive to eliminate I/O as
performance bottleneck.

\begin{table}[t]
	\centering
	\begin{tabular}{lcc}
	\hline
	\textbf{Method} & \textbf{Analysis window} & \textbf{Run time} \\
	\hline
	Churn & Two consensuses & $\sim$0.16s \\
	Neighbor search & One consensus & $\sim$1.6s \\
	Fingerprint & One month & $\sim$58s \\
	Uptimes & One month & $\sim$145s \\
	\hline
	\end{tabular}
	\caption{The computational cost of our analysis techniques.}
	\label{tab:exp-deployment}
\end{table}

The table columns contain, from left to right, our analysis technique, the
technique's analysis window, and how long it takes to compute its output.
Network churn calculation is very fast; it takes as input only two consensus
files and can easily be run for every new network consensus.  Nearest-neighbor
search takes approximately 1.6 seconds for a single consensus counting 6,942
relays.  Fingerprint and uptime analysis for one month worth of consensuses
takes approximately one and two minutes, respectively---easy to invoke daily, or
even several times a day.

\section{Discussion}\label{sec:discussion}
After having used \sys in practice for several months, we elaborate on both our
operational experience and the shortcomings we encountered.

\subsection{Operational experience}
\label{sec:operational}
Our practical work with \sys taught us that detecting Sybils frequently requires
manual work; for example, comparing a new Sybil group with a previously
disclosed one, sending decoy traffic over Sybils, or sorting and comparing
information in their descriptors.  It is difficult to predict all kinds of
manual analyses that might be necessary in the future, which is why we designed
\sys to be highly interoperable with Unix command line tools~\cite{Pike1983a}.
Its CSV-formatted output can easily be piped into tools such as sed, awk, and
grep.  We found that compact text output was significantly easier to process,
both for plotting results and for manual analysis.  We also found that \sys can
serve as valuable tool to better understand the Tor network and monitor its
reliability.  Our techniques can disclose network consensus issues and
illustrate the wide diversity of Tor relays, providing empirical data that can
support future network design decisions.

We are also working with The Tor Project on incorporating our techniques in
Tor Metrics~\cite{metrics}, a web site that contains network visualizations,
which are frequented by numerous volunteers that sometimes report anomalies.
By incorporating our techniques, we hope to benefit from ``crowd-sourced'' Sybil
detection.

\subsection{Limitations}
\label{sec:limitations}
In Section~\ref{sec:threat_model} we argued that we are unable to prevent all
Sybil attacks.  An adversary unconstrained by time and money can add an
unlimited number of Sybils to the network.  Indeed, Table~\ref{tab:sybils}
contains six Sybil groups that \sys was unable to detect.  Exitmap, however, was
able to expose these Sybils, which emphasizes the importance of having diverse
and complementary analysis techniques to raise the bar for adversaries.  By
characterizing past attacks and documenting the evolution of recurring attacks,
we can adapt our techniques, allowing for the bar to be raised even further.
However, this arms race is unlikely to end, barring fundamental changes in how
Tor relays are operated.  Given that attackers can stay under our radar, our
results represent a lower bound because we might have missed Sybil groups.

Finally, \sys is unable to ascertain the purpose of a Sybil attack.  While the
purpose is frequently obvious, Table~\ref{tab:sybils} contains several Sybil
groups that we could not classify.  In such cases, it is difficult for The Tor
Project to make a call and decide if Sybils should be removed from the network.
Keeping them runs the risk of exposing users to an unknown attack, but removing
them deprives the network of bandwidth.  Often, additional context is helpful in
making a call.  For example, Sybils that are (\emph{i}) operated in
``bulletproof'' ASes~\cite{Konte2015a}, (\emph{ii}) show signs of not running
the Tor reference implementation, or (\emph{iii}) spoof information in their
router descriptor all suggest malicious intent.  In the end, Sybil groups have
to be evaluated case by case, and the advantages and disadvantages of blocking
them have to be considered.

\section{Conclusion}
\label{sec:conclusion}
We presented \sys, a novel system that uses diverse analysis techniques to
expose Sybils in the Tor network.  Equipped with this tool, we set out to
analyze nine years of The Tor Project's archived network data.  We discovered
numerous Sybil groups, twenty of which we present in this work.  By analyzing
these Sybil groups \sys discovered, we found that (\emph{i}) Sybil relays are
frequently configured very similarly, and join and leave the network
simultaneously; (\emph{ii}) attackers differ greatly in their technical
sophistication; and (\emph{iii}) our techniques are not only useful for spotting
Sybils, but turn out to be a handy analytics tool to monitor and better
understand the Tor network.  Given the lack of a central identity-verifying
authority, it is always possible for well-executed Sybil attacks to stay under
our radar, but we found that a complementary set of techniques can go a long way
towards finding malicious Sybils, making the Tor network more secure and
trustworthy for its users.

Both code and data for this work are available online at
\url{https://nymity.ch/sybilhunting/}.

\section*{Acknowledgments}
This research was supported in part by the Center for Information Technology
Policy at Princeton University.  We want to thank Stefan Lindskog and the Tor
developers for helpful feedback.

\newpage

\printbibliography

\appendix

\section{Exposed malicious exit relays}
\label{sec:malicious-relays}
Table~\ref{tab:exitmap-dataset} provides an overview of our second dataset, 251
bad exit relays that we discovered between August 2014 and January 2016.  We
believe that all single relays in the dataset were isolated incidents while sets
of relays constituted Sybil groups.  Sybil groups marked with the symbols
$\dagger$, $\ddagger$, and $\mathsection$ were run by the same attacker.

\rowcolors{1}{}{gray!10}

\begin{table*}
\small
\centering
\begin{tabularx}{\textwidth}{r c X}
\hline
\textbf{Discovery} & \textbf{\# of relays} & \textbf{Attack description} \\
\hline
Aug 2014 & 1 & The relay injected JavaScript into returned HTML.  The script
embedded another script from the domain fluxx.crazytall.com---not clearly
malicious, but suspicious. \\

Aug 2014 & 1 & The relay injected JavaScript into returned HTML.  The script
embedded two other scripts, jquery.js from the official jQuery domain, and
clr.js from adobe.flashdst.com.  Again, this was not necessarily malicious, but
suspicious. \\

Sep 2014 & 1 & The exit relay routed traffic back into the Tor network, i.e., we
observed traffic that was supposed to exit from relay $A$, but came from relay
$B$.  The system presented by Ling et al. behaves the same~\cite{Ling2015a};
the authors proposed to run intrusion detection systems on Tor traffic by
setting up an exit relay that runs an NIDS system, and routes the traffic back
into the Tor network after having inspected the traffic. \\

Oct 2014 & 1 & The relay injected JavaScript into returned HTML. \\

Oct 2014 & 1 & The relay ran the MitM tool sslstrip~\cite{sslstrip}, rewriting
HTTPS links to unencrypted HTTP links in returned HTML. \\

Oct 2014 & 1 & Same as above. \\

Jan 2015 & 23$\ddagger$ & Blockchain.info's web server redirects its
users from HTTP to HTTPS.  These relays tampered with blockchain.info's redirect
and returned unprotected HTTP instead---presumably to sniff login credentials. \\

Jan 2015 & 1 & The relay used OpenDNS as DNS resolver and had the web site
category ``proxy/anonymizer'' blocked, resulting in several inaccessible web
sites, including torproject.org. \\

Feb 2015 & 1 & The relay injected a script that attempted to load a resource
from the now inaccessible torclick.net.  Curiously, torclick.net's front page
said ``We place your advertising materials on all websites online.  Your ads
will be seen only for anonymous network TOR [sic] users.  Now it is about 3
million users. The number of users is always growing.'' \\

Feb 2015 & 17$\ddagger$ & Again, these relays tampered with HTTP redirects of
Bitcoin web sites.  Interestingly, the attack became more sophisticated; these
relays would only target connections whose HTTP headers resembled Tor Browser.
\\

Mar 2015 & 18$\ddagger$ & Same as above. \\

Mar 2015 & 1 & The relay injected JavaScript and an iframe into the returned
HTML.  The injected content was not clearly malicious, but suspicious. \\

Apr 2015 & 70$\dagger$ & These exit relays transparently rewrote onion domains
in returned HTML to an impersonation domain.  The impersonation domain looked
identical to the original, but had different Bitcoin addresses.  We believe that
this was attempt to trick Tor users into sending Bitcoin transactions to
phishing addresses. \\

Jun 2015 & 55$\dagger$ & Same as above. \\

Aug 2015 & 4$\dagger$ & Same as above. \\

Sep 2015 & 1 & The relay injected an iframe into returned HTML that would load
content that made the user's browser participate in some kind of mining
activity. \\

Nov 2015 & 1 & The relay ran the MitM tool sslstrip. \\

Nov 2015 & 8$\dagger$ & Same as the relays marked with a $\dagger$. \\

Dec 2015 & 1$\mathsection$ & The relay ran the MitM tool sslstrip. \\

Dec 2015 & 1$\mathsection$ & Same as above. \\

Jan 2016 & 43$\dagger$ & Same as the relays marked with a $\dagger$. \\
\hline
\end{tabularx}
\caption{An overview of our second dataset, 251 malicious exit relays that we
discovered using exitmap.  We believe that Sybil groups marked with an
$\dagger$, $\mathsection$, and $\ddagger$ were run by the same adversary.}
\label{tab:exitmap-dataset}
\end{table*}

\section{Supporting diagrams}
\begin{figure}[h]
	\centering
	\includegraphics[width=0.8\linewidth]{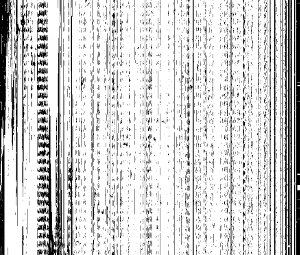}
	\caption{Uptimes for the ``default'' Sybil group for October 2015.  Many
	relays exhibit a diurnal pattern, suggesting that the relays were powered
	off regularly.}
	\label{fig:default-sybils-uptime}
\end{figure}

Figure~\ref{fig:default-sybils-uptime} shows the uptime matrix for the
``default'' Sybil group for October 2015.  Matrix rows represent consensuses and
columns represent relays.  As a result, a single pixel shows if a given relay
was online (black pixel) or offline (white pixel) in a given consensus.  The
matrix shows that many relays exhibit a diurnal uptime pattern.

\end{document}